\documentclass[12pt,a4paper]{article}
\usepackage{graphicx}
\newcommand{\Li}[2]{{\mbox{Li}}_{#1}\left(#2\right)}
\newcommand{\snp}[2]{{\mbox{S}}_{#1}\left(#2\right)}

\begin{document}

\begin{flushright}
 {Freiburg-THEP 99/10}\\
 {hep-ph/9909506}\\
 {September 1999}
\end{flushright}

\vspace{2 cm}

\begin{center}
 {\large
Non-planar massless two-loop Feynman diagrams with four on-shell legs
 }\\[3ex]
J.B. Tausk\footnote{tausk@physik.uni-freiburg.de}
\\[3ex]

Fakult{\"a}t f{\"u}r Physik,
Albert-Ludwigs-Universit{\"a}t Freiburg,\\
Hermann-Herder-Stra{\ss}e 3,
D-79104 Freiburg, Germany

\vspace{1 cm}

\begin{abstract}
The non-planar Feynman diagram with seven massless, scalar propagators and
four on-shell legs (the crossed double box) is calculated analytically in
dimensional regularization. The non-planar diagram with six propagators
is also discussed.
\end{abstract}

\end{center}

\vspace{1ex}
{\bf 1. Introduction}\\
Knowledge of two-loop massless Feynman diagrams with four
massless external legs is one of the key ingredients required for
a next-to-next-to-leading order calculation of 2 jet production
rates in hadron-hadron collisions \cite{GloverMoriond98}, as well
as for other processes, including Bhabha scattering in the high energy
limit. Among those diagrams, the most complicated are the planar and
non-planar double box diagrams, both of which contain seven propagators.
Recently, significant progress in this field has been made by Smirnov, who
has derived an analytical formula in terms of polylogarithms
for the planar double box \cite{Smirnovdoublebox} (see also
ref.~\cite{Ussyukina}). More simple
planar diagrams, with less than seven propagators, are discussed in
refs.~\cite{SmirnovVeretin99,AGO99ii}. Results for non-planar diagrams
have until now been missing.
\begin{center}
\includegraphics[height=1.5cm]{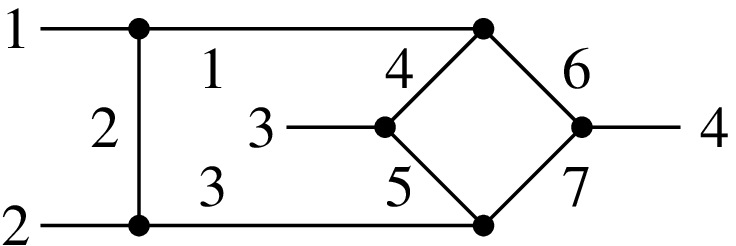}
\hspace{2cm}
\includegraphics[height=1.5cm]{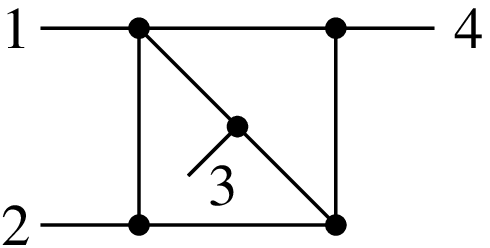}
\end{center}

The ultraviolet divergences of the planar and non-planar double box
diagrams in spacetime dimensions $d=7$, $d=9$ and $d=11$ were calculated
in ref.~\cite{BernDDPRozowsky98}. In $d=4$, the diagrams are ultraviolet
finite, but they have a complicated structure of collinear and infrared
divergences. In dimensional regularization, with $d=4-2\epsilon$, the
divergences appear as poles $1/\epsilon^j$, with $j=1,2,3,4$. As has been
demonstrated by Smirnov in the planar case, the method of Mellin-Barnes
contour integrals is a very natural and convenient technique to isolate
these singular contributions. The main purpose of this paper is to
calculate the non-planar seven propagator diagram,
\begin{equation}
\label{eq:XI}
B_7^{NP} =
\int \int \mbox{d}^d k  \; \mbox{d}^d l \;
\frac{1}{(k+l+p_1+p_2)^2 (k+l+p_2)^2 (k+l)^2 (k-p_3)^2 k^2 (l-p_4)^2 l^2 }
\, ,
\end{equation}
using a similar Mellin-Barnes approach. The external momenta $p_1\ldots p_4$
are lightlike ($p_i^2=0$) and we use Mandelstam variables defined by
$s=(p_1+p_2)^2$, $t=(p_1+p_3)^2$ and $u=(p_1+p_4)^2$.

Starting from a Mellin-Barnes representation, which we derive in section 2,
we obtain analytical formulas in terms of logarithms and polylogarithms
for the seven propagator diagram $B_7^{NP}$, which we present in section 5,
by a series of steps that are explained in sections 3 and 4. Then, in
section 6, we briefly discuss the non-planar diagram with six propagators.
Some final comments are made in section 7.

\vspace{1ex}
{\bf 2. A Mellin-Barnes representation}\\
Our main tool will be the Mellin-Barnes representation of a power of
a sum as a contour integral,
\begin{equation}
\label{eq:MB2}
(A_1+A_2)^{-\nu} =
 \frac{1}{2 \pi i}
 \int_{-i\infty}^{i\infty} \mbox{d} \sigma
\, A_1^{\sigma} A_2^{-\nu-\sigma}
 \frac{\Gamma(-\sigma) \Gamma(\nu+\sigma)}
      {\Gamma(\nu)} \, ,
\end{equation}
where the integration contour separates the poles of $\Gamma(-\sigma)$
from those of $\Gamma(\nu+\sigma)$, and $A_{1,2}$ are complex numbers
such that $|\arg(A_1)-\arg(A_2)|<\pi$.  The Mellin-Barnes representation
of a sum of several terms, $(A_1+\ldots+A_n)^{-\nu}$, is obtained easily
by iteration of the basic formula~(\ref{eq:MB2}). Another formula that
we will need is Barnes's first lemma (see, e.g. \cite{Slater}):
\begin{eqnarray}
\label{eq:BL1}
 \frac{1}{2 \pi i}
 \int_{-i\infty}^{i\infty} \mbox{d} \sigma
\,\Gamma({\sigma_1+\sigma})
\,\Gamma({\sigma_2+\sigma})
\,\Gamma({\sigma_3-\sigma})
\,\Gamma({\sigma_4-\sigma})
\nonumber \\
= \frac{
  \Gamma({\sigma_1+\sigma_3})
\,\Gamma({\sigma_1+\sigma_4})
\,\Gamma({\sigma_2+\sigma_3})
\,\Gamma({\sigma_2+\sigma_4})
  }{\Gamma({\sigma_1+\sigma_2+\sigma_3+\sigma_4})}
\, .
\end{eqnarray}
Again, the contour should separate the increasing series of poles
(of $\Gamma({\sigma_{3,4}-\sigma})$) from the decreasing ones
(of $\Gamma({\sigma_{1,2}+\sigma})$).

Introducing Feynman parameters in the standard way, the non-planar double
box diagram can be written as \cite{BernDDPRozowsky98}
\begin{eqnarray}
\label{eq:XIFP}
B_7^{NP} & = & (-1)^7 {(i \pi^{d/2})}^2 \Gamma(7-d)
\int_0^\infty \mbox{d} x_1 \ldots \mbox{d} x_7
\, \delta(1-x_1 \ldots -x_7)
\, D^{7-\frac{3d}{2}} C^{d-7} 
\\
D & = & (x_1+x_2+x_3)(x_4+x_5+x_6+x_7) + (x_4+x_5)(x_6+x_7)
\\
C & = & -t \, x_2 x_5 x_6 - u \, x_2 x_4 x_7
        -s \, x_1 x_5 x_7 - s \, x_3 x_4 x_6
\nonumber \\
  &   & -s \, x_1 x_3 (x_4+x_5+x_6+x_7) \; .
\end{eqnarray}
According to the causal $i\epsilon$-prescription, a small imaginary
part should be added to $C$. We do this by giving $s$, $t$ and $u$
positive imaginary parts. At this stage, we consider $s$, $t$ and $u$
to be three independent variables. At the end of the calculation,
after we have taken the limit where $s$, $t$ and $u$ become real, one
variable can be eliminated by using the relationship $s+t+u=0$.

The next step is to represent the factor $C^{d-7}$ as a product of
five more simple factors, at the expense of introducing four
Mellin-Barnes parameters:
\begin{eqnarray}
\label{eq:Cdecomp}
 \Gamma(7-d) \, C^{d-7} & = &
 \frac{1}{(2 \pi i)^4}
 \int_{-i\infty}^{i\infty} \mbox{d} \xi_1
 \int_{-i\infty}^{i\infty} \mbox{d} \xi_2
 \int_{-i\infty}^{i\infty} \mbox{d} \alpha
 \int_{-i\infty}^{i\infty} \mbox{d} \beta
\nonumber \\ &&
\times
\Gamma(-\xi_1) \,\Gamma(-\xi_2) \,\Gamma(-\alpha) \,\Gamma(-\beta)
\,\Gamma(7-d+\xi_1+\xi_2+\alpha+\beta)
\nonumber \\ &&
\times
\,\left[-t \, x_2 x_5 x_6\right]^{\xi_1}
  \left[-u \, x_2 x_4 x_7\right]^{\xi_2}
  \left[-s \, x_1 x_5 x_7\right]^{\alpha}
  \left[-s \, x_3 x_4 x_6\right]^{\beta}
\nonumber \\ &&
\times
\left[-s \, x_1 x_3 (x_4+x_5+x_6+x_7)\right]^{d-7-\xi_1-\xi_2-\alpha-\beta}
\, .
\end{eqnarray}
Now the Feynman parameter integrals can be done in terms of
$\Gamma$-functions, and we obtain the following Mellin-Barnes
representation for $B_7^{NP}$:
\begin{equation}
B_7^{NP} = \frac{\pi^d \,\Gamma^2(-\epsilon)}
     {\Gamma(-2\epsilon) \,\Gamma(-1-3\epsilon)} K \, ,
\end{equation}
\begin{eqnarray}
\label{eq:XIMB}
K & = &
 \frac{1}{(2 \pi i)^4}
 \int_{-i\infty}^{i\infty} \mbox{d} \xi_1
 \int_{-i\infty}^{i\infty} \mbox{d} \xi_2
 \int_{-i\infty}^{i\infty} \mbox{d} \alpha
 \int_{-i\infty}^{i\infty} \mbox{d} \beta
\nonumber \\ &&
\times
\left(-s \right)^{-3-2\epsilon-\xi_1-\xi_2}
\left(-t \right)^{\xi_1}
\left(-u \right)^{\xi_2}
\frac{ \Gamma_1 \Gamma_2 \Gamma_3 \Gamma_4
\Gamma_5 \Gamma_6 \Gamma_7 \Gamma_8 \Gamma_9
\Gamma_{10} \Gamma_{11} \Gamma_{12} \Gamma_{13} }
{ \Gamma_{14}^2 }
\, ,
\end{eqnarray}
where

\begin{equation}
\label{eq:gammalist}
\begin{array}{ll}
 \Gamma_{1}  =  \Gamma(3+2 \epsilon + \alpha + \beta + \xi_1 + \xi_2)
&\Gamma_{8}  =  \Gamma(-2-2 \epsilon - \alpha - \xi_1 - \xi_2)
\\
 \Gamma_{2}  =  \Gamma(2 + \epsilon + \alpha + \beta + \xi_1 + \xi_2)
&\Gamma_{9}  =  \Gamma(-2-2 \epsilon - \beta - \xi_1 - \xi_2)
\\
 \Gamma_{3}  =  \Gamma(1 + \xi_1 + \xi_2)
&\Gamma_{10}  =  \Gamma(-\alpha)
\\
 \Gamma_{4}  =  \Gamma(1 + \alpha + \xi_1)
&\Gamma_{11}  =  \Gamma(-\beta)
\\
 \Gamma_{5}  =  \Gamma(1 + \alpha + \xi_2)
&\Gamma_{12}  =  \Gamma(-\xi_1)
\\
 \Gamma_{6}  =  \Gamma(1 + \beta + \xi_1)
&\Gamma_{13}  =  \Gamma(-\xi_2)
\\
 \Gamma_{7}  =  \Gamma(1 + \beta + \xi_2)
&\Gamma_{14}  =  \Gamma(2 + \alpha + \beta + \xi_1 + \xi_2)
\, .
\end{array}
\end{equation}

In order to keep things as simple as possible, we choose the contour
for each of the variables $\xi_1$, $\xi_2$, $\alpha$ and $\beta$ to be a
straight line parallel to the imaginary axis, with a real part that does
not depend on any of the other variables. Eq.~(\ref{eq:XIMB}) is certainly
valid if the real parts of the contours are chosen in such a way that
the real parts of the arguments of $\Gamma_{1} \ldots \Gamma_{13}$ are
all positive. It is possible to satisfy these conditions simultaneously
when $-3/2<\Re(\epsilon)<-1/3$. For example, if $\epsilon\approx-1/2$,
choosing
$\Re(\xi_1)=\Re(\xi_2)=-5/12$, $\Re(\alpha)=\Re(\beta)=-1/4$, satifies
all the conditions. Obviously, there are infinitely many other, equally
suitable possibilities, but for the sake of definiteness, we will stick
to these contours.

\vspace{1ex}
{\bf 3. Analytic continuation in $\epsilon$}\\
Having derived the Mellin-Barnes representation~(\ref{eq:XIMB})
near $\epsilon=-1/2$, we must now perform an analytic continuation in
$\epsilon$ to the region around $\epsilon=0$. To do so, we follow the
poles of the $\Gamma$-functions as we increase $\epsilon$. Whenever
a pole crosses one of the integration contours, which we keep fixed,
we add its residue in the corresponding variable to the right hand
side of eq.~(\ref{eq:XIMB}). This residue term still contains poles
in its remaining variables which can cross other integration contours
as $\epsilon$ is increased further. They are treated in the same way.
We end up with a large collection of residue terms, plus the original
fourfold integral.
Some of the residue terms contain factors that diverge at $\epsilon=0$.
However, these singular factors do not depend on $\xi_1$, $\xi_2$,
$\alpha$ or $\beta$. The remaining integrals can simply be expanded
in $\epsilon$ under the integral sign because our contours do not go
through any poles at $\epsilon=0$.
(There are never problems with convergence at the limits
$\xi_1,\xi_2,\alpha,\beta\rightarrow\pm i\infty$).

Below, we will give some examples of the residue terms that appear. But
first, let us see what happens when we substitute $\epsilon=0$ in the
integral~(\ref{eq:XIMB}). The variables $\alpha$ and $\beta$ are coupled
via $\Gamma_1$, $\Gamma_2$ and $\Gamma_{14}$,
but these $\Gamma$-functions collapse into
\begin{equation}
\label{eq:collapse}
\Gamma_1 \Gamma_2 / \Gamma_{14}^2 
= 2+\alpha+\beta+\xi_1+\xi_2
= (1+\alpha+\xi_1)+(1+\beta+\xi_2)
\, .
\end{equation}
Consider the first term. The factor $(1+\alpha+\xi_1)$ can be
absorbed into $\Gamma_{4}$, and then both the $\alpha$ and
the $\beta$ integrals can be performed using Barnes's first
lemma. Moreover, the $\beta$ integral gives {\em zero:}
\begin{eqnarray}
\label{eq:zero}
\frac{1}{2 \pi i} 
 \int_{-i\infty}^{i\infty} \mbox{d} \beta
\,\Gamma(1+\beta+\xi_1)\,\Gamma(1+\beta+\xi_2)
\,\Gamma(-2-\beta-\xi_1-\xi_2) \,\Gamma(-\beta)
\nonumber \\
= \frac{\Gamma(-1-\xi_1)\,\Gamma(-1-\xi_2)
\,\Gamma(1+\xi_1)\,\Gamma(1+\xi_2)}{\Gamma(0)} = 0
\, .
\end{eqnarray}
(Here, we are taking a contour that separates the increasing and
decreasing series of poles as required by Barnes's lemma, not a
straight line).  Analogously, the second term on the right hand side of
eq.~(\ref{eq:collapse}) vanishes when integrated over $\alpha$.

This cancellation is very nice because it means that if we only need
the non-planar box up to the constant term in $\epsilon$, once we have
done the $\alpha$ and $\beta$ integrals, we do not need to calculate
any double integrals of the form
\begin{equation}
\label{eq:double}
 \int_{-i\infty}^{i\infty} \mbox{d} \xi_1
 \int_{-i\infty}^{i\infty} \mbox{d} \xi_2
\left(-s \right)^{-3-\xi_1-\xi_2} 
\left(-t \right)^{\xi_1} 
\left(-u \right)^{\xi_2}
\ldots \, .
\end{equation}
Only terms in which a residue has been taken in either $\xi_1$ or $\xi_2$,
or in both, are ever needed. Such terms, whose dependence on at least
one of the variables $s$, $t$ and $u$ is trivial, are much more simple
than the double integrals~(\ref{eq:double}).

We now return to the task of collecting the various residue
contributions. The first poles to cross contours are at
$\alpha=-2-2 \epsilon - \xi_1 - \xi_2$
and
$\beta=-2-2 \epsilon - \xi_1 - \xi_2$,
coming from $\Gamma_{8}$ and $\Gamma_{9}$, respectively. (The poles
of $\Gamma_{1}$ and $\Gamma_{2}$ are harmless because they depend
on $\epsilon$ with the opposite sign, and therefore move away from
the contours when $\Re(\epsilon)$ increases). To account for the
residues of these two poles, we replace
$K \rightarrow K+K_8+K_9+K_{89}$, where
\begin{eqnarray}
K_{8} & = &
  \frac{1}{(2 \pi i)^3}
 \int_{-i\infty}^{i\infty} \mbox{d} \xi_1
 \int_{-i\infty}^{i\infty} \mbox{d} \xi_2
 \int_{-i\infty}^{i\infty} \mbox{d} \beta
\left(-s \right)^{-3-2\epsilon-\xi_1-\xi_2}
\left(-t \right)^{\xi_1}
\left(-u \right)^{\xi_2}
\nonumber \\ &&
\times
\left.
\frac{ \Gamma_1 \Gamma_2 \Gamma_3 \Gamma_4
\Gamma_5 \Gamma_6 \Gamma_7 \;\; \Gamma_9
\Gamma_{10} \Gamma_{11} \Gamma_{12} \Gamma_{13} }
{ \Gamma_{14}^2 }
\right|_{\alpha=-2-2 \epsilon - \xi_1 - \xi_2}
\, ,
\end{eqnarray}
$K_9$ is a similar term, where the residue in $\beta$
is taken, and
\begin{eqnarray}
K_{89} & = &
  \frac{1}{(2 \pi i)^2}
 \int_{-i\infty}^{i\infty} \mbox{d} \xi_1
 \int_{-i\infty}^{i\infty} \mbox{d} \xi_2
\left(-s \right)^{-3-2\epsilon-\xi_1-\xi_2}
\left(-t \right)^{\xi_1}
\left(-u \right)^{\xi_2}
\nonumber \\ &&
\times
\left. 
\frac{ \Gamma_1 \Gamma_2 \Gamma_3 \Gamma_4
\Gamma_5 \Gamma_6 \Gamma_7 \;\; \;\;
\Gamma_{10} \Gamma_{11} \Gamma_{12} \Gamma_{13} }
{ \Gamma_{14}^2 }
\right|_{\alpha=\beta=-2-2 \epsilon - \xi_1 - \xi_2}
\, .
\end{eqnarray}
Let us concentrate on $K_8$. Note that in eliminating $\alpha$, we have changed
the $\epsilon$-dependence of the remaining $\Gamma$-functions.
As a consequence, the first poles of
$\Gamma_{2}=\Gamma(\beta-\epsilon)$,
$\Gamma_{4}=\Gamma(-1-2\epsilon-\xi_2)$
and
$\Gamma_{5}=\Gamma(-1-2\epsilon-\xi_1)$
cross the $\beta$, $\xi_2$ and $\xi_1$ contours, respectively. These crossings
are taken into account by replacing
$K_8 \rightarrow K_8+K_{82}+K_{84}+K_{85}+K_{842}+K_{852}+K_{845}+K_{8452}$.
In a number of these terms, there are still more poles that cross
contours before $\epsilon$ finally reaches $0$. They are dealt with
by making the following further replacements:
$K_{82} \rightarrow K_{82}+K_{829'}$,
$K_{84} \rightarrow K_{84}+K_{843}$,
$K_{842} \rightarrow K_{842}+K_{8423}$,
$K_{85} \rightarrow K_{85}+K_{853}$,
$K_{852} \rightarrow K_{852}+K_{8523}$,
and
$K_{845} \rightarrow K_{845}-K_{8459}$.
In each term, the subscripts indicate the $\Gamma$-functions
of which the residue of the first pole has been taken. An exception
to this is $K_{829'}$, where the prime means that we have taken
the residue of the second pole of $\Gamma_9$ (i.e., where the
argument equals -1) instead of the first pole. The last term,
$K_{8459}$, gets a relative minus sign because the pole in
$\Gamma_9$ crosses the $\beta$ contour backwards.\footnote{
Eventually, $K_{8459}$ cancels against a descendant of $K_{89}$.}

Due to the symmetry of the problem in $\alpha$ and $\beta$, there is a one
to one correspondence between the descendants of $K_9$ and those of $K_8$,
and their contributions to the final result are the same. The analysis
of $K_{89}$ runs along similar lines. Here, some poles are second order,
with residues involving derivatives of $\Gamma$-functions.

\vspace{1ex}
{\bf 4. Evaluating the $K$'s}\\
In all, we find 39 contributions. (This number depends on the
contours one uses. Had we chosen different ones,
there would have been other poles crossing contours and we would have
obtained a different, but equivalent, collection of residue terms).
The simplest contributions are those where a residue in four variables has
been taken, e.g.
\begin{eqnarray}
\label{eq:K8452}
K_{8452} =
\left.
\left(-s \right)^{-3-2\epsilon-\xi_1-\xi_2}
\left(-t \right)^{\xi_1}
\left(-u \right)^{\xi_2}
\frac{ \Gamma_1 \Gamma_3
\Gamma_6 \Gamma_7 \Gamma_9
\Gamma_{10} \Gamma_{11} \Gamma_{12} \Gamma_{13} }
{ \Gamma_{14}^2 }
\right|_{\begin{array}{l}
\xi_1=\xi_2=-1-2\epsilon\\
\alpha=2\epsilon, \; \beta=\epsilon
\end{array}}
\nonumber \\ =
\left(-s \right)^{-1+2\epsilon}
\left(-t \right)^{-1-2\epsilon}
\left(-u \right)^{-1-2\epsilon}
\,\Gamma(1+\epsilon) \,\Gamma(-1-4\epsilon) \,\Gamma(\epsilon)
\,\Gamma(-2\epsilon) \,\Gamma(-\epsilon) \,\Gamma^2(1+2\epsilon)
\, .
\end{eqnarray}
Note that in this particular case, the $\epsilon$ expansion
starts off with a $1/\epsilon^4$ pole.

In several terms, there is one integral left. For instance, $K_{843}$,
which we get by taking residues at
$\alpha=-2-2\epsilon-\xi_1-\xi_2$,
$\xi_2=-1-2\epsilon$
and
$\xi_1=2\epsilon$,
is given by
\begin{eqnarray}
&& K_{843} =
\left(-s \right)^{-2-2\epsilon}
\left(-t \right)^{2\epsilon}
\left(-u \right)^{-1-2\epsilon}
\,\Gamma(-1-4\epsilon) \,\Gamma^2(1+2\epsilon) \,\Gamma(-2\epsilon)
\nonumber \\
&& \times
\frac{1}{2 \pi i} 
 \int_{-i\infty}^{i\infty} \mbox{d} \beta
\,\frac{\Gamma(1+\beta) \,\Gamma(\beta-\epsilon)
\,\Gamma(1+2\epsilon+\beta) \,\Gamma(-1-2\epsilon-\beta)
\,\Gamma(-\beta)}{\Gamma(\beta-2\epsilon)}
\, .
\end{eqnarray}
To calculate the $\beta$ integral, we expand the integrand up to the
second order in $\epsilon$. Then, closing the contour to the left
or right and summing the residues of the poles inside it, we obtain
harmonic series which can be expressed in terms of $\zeta(2)$, $\zeta(3)$
and $\zeta(4)$.

An example in which the final integral still involves powers of
Mandelstam variables is
\begin{equation}
\label{eq:k842}
K_{842} =
\frac{-1}{2 \pi i\, u} 
 \int_{-i\infty}^{i\infty} \mbox{d} \xi_1
\left(-s \right)^{-2-\xi_1}
\left(-t \right)^{\xi_1}
\,\frac{\Gamma^3(1+\xi_1)\,\Gamma^3(-\xi_1)}
{\xi_1 (1+\xi_1)^2}
\, ,
\end{equation}
where we have already set $\epsilon=0$. This integral can also be
calculated by closing the contour, and the result can be expressed
in terms of polylogarithms $\mbox{Li}_n$ \cite{Lewin} and Nielsen's
generalized polylogarithms $\mbox{S}_{n,p}$
\cite{Nielsen,KolbigMignacoRemiddi70}, which are defined by
\begin{equation}
\snp{n,p}{z} = \frac{(-1)^{n-1+p}}{(n-1)!\,p!}
\int_0^1 \mbox{d} t \; \frac{ \log^{n-1}(t) \log^p(1-zt) }{t} \, .
\end{equation}
The cases $\mbox{S}_{1,2}$, $\mbox{S}_{1,3}$ and $\mbox{S}_{2,2}$ which
occur in this paper can all be written as combinations of $\mbox{Li}_2$,
$\mbox{Li}_3$ and $\mbox{Li}_4$.

In the term $K_{84}$, we perform the first integration by Barnes's
lemma (in a slightly modified form, see \cite{Smirnovdoublebox}):
\begin{eqnarray}
K_{84} & = & 
  \frac{1}{(2 \pi i)^2\,u}
 \int_{-i\infty}^{i\infty} \mbox{d} \xi_1
\left(-s \right)^{-2-\xi_1}
\left(-t \right)^{\xi_1}
\,\frac{\Gamma^2(1+\xi_1)\,\Gamma^2(-\xi_1)}
{\xi_1 (1+\xi_1)}
\nonumber \\ && \times
 \int_{-i\infty}^{i\infty} \mbox{d} \beta
\,\Gamma(1+\beta) \,\Gamma(1+\beta+\xi_1)
\,\Gamma(-1-\beta-\xi_1)\,\Gamma(-\beta)
\nonumber \\ & = &
\frac{1}{2 \pi i\, u}
 \int_{-i\infty}^{i\infty} \mbox{d} \xi_1
\left(-s \right)^{-2-\xi_1}
\left(-t \right)^{\xi_1}
\,\frac{\Gamma^3(1+\xi_1)\,\Gamma^3(-\xi_1)}
{\xi_1 (1+\xi_1)}
\left( \psi(1+\xi_1) - \psi(1) \right) \,.
\end{eqnarray}
The $\xi_1$ integral can be then be done by closing the contour,
as in the previous example. Three other $K$'s are related to $K_{84}$
by $\alpha\leftrightarrow\beta$ and/or $\xi_1\leftrightarrow\xi_2$.

Finally, there are four terms\footnote{We are leaving out $K_{82}$
and $K_{92}$, which contain a factor of $\Gamma(-\epsilon)$ in their
denominator and are therefore already of order $\epsilon$.} in which
both the $\xi_1$ and the $\xi_2$ integrals still survive, namely $K$,
$K_8$, $K_9$ and $K_{89}$. However, by the mechanism explained above
(see eq.~(\ref{eq:zero})), these terms cancel against each other at
$\epsilon=0$ after integration over $\alpha$ and $\beta$.

\vspace{1ex}
{\bf 5. Result for the seven propagator diagram}\\
Adding together all the contributions discussed in the previous section,
we obtain an expression for $B_7^{NP}$, up to the constant term in
$\epsilon$, in terms of (poly)logarithms of (ratios of) $s$, $t$ and $u$,
which are, at this point, still independent variables. In particular,
the result is also valid when $s$, $t$ and $u$ are all negative. This
region is, of course, unphysical, but it is nevertheless interesting to
consider because it provides us with an opportunity to check our result
against a calculation by Binoth and Heinrich \cite{HeinrichBinoth}. They
have obtained results for the negative $s$, $t$, $u$ region by extracting
the divergences from the Feynman parametric integral~(\ref{eq:XIFP})
and then performing a multi-dimensional numerical integration.

At the symmetric point, $s=t=u=-1$, we find
\begin{eqnarray}
\label{eq:B7-111}
B_7^{NP} & = & \pi^d \,\Gamma(3+2\epsilon) \left\{
\frac{7}{4\epsilon^4} - \frac{3}{\epsilon^3}
+\frac{1}{\epsilon^2}\left(\frac{-7}{2}-\frac{47}{24}\pi^2\right)
+\frac{1}{\epsilon}\left(
  \frac{105}{2}+\frac{37}{4}\pi^2-\frac{89}{4}\zeta(3)
\right.\right. \nonumber \\ && \left.
\hspace{5em}
    -\frac{1}{2}\pi^2\log(2)\right)
 -\frac{589}{2}-\frac{112}{3}\pi^2-\frac{443}{288}\pi^4
 +\frac{555}{4}\zeta(3)
\nonumber \\ && \left.
\hspace{5em}
 +\frac{75}{2}\pi^2\log(2)-\frac{11}{3}\pi^2\log^2(2)
 +\frac{11}{12}\log^4(2) + 22\,\Li{4}{1/2}
 \right\}
\nonumber \\ & = &  \pi^d \,\Gamma(3+2\epsilon) \left\{
 \frac{1.75}{\epsilon^4}
 -\frac{3}{\epsilon^3}
 -\frac{22.828}{\epsilon^2}
 +\frac{113.63}{\epsilon}
 -395.26 \right\} \, .
\end{eqnarray}
The coefficients in (\ref{eq:B7-111}) agree with the results obtained
by the numerical method \cite{HeinrichBinoth}.  We also find good
agreement at the asymmetric point $(s,t,u)=(-1,-2,-3)$. In both cases
the precision of the numerical method ranges from about 1 per mille for
the $1/\epsilon^4$-term to 1 per cent for the constant term.

Since the complete formula for general $s$, $t$ and $u$ is very long
and not relevant for physical applications anyway, we do not present it
here. Instead, we specialize to the physically relevant case
where $s+t+u=0$. 
We also use transformation formulas for the
arguments of the polylogarithms in order to avoid ending up
on a cut when we make $s$, $t$, and $u$ real. Depending on which
legs correspond to incoming and outgoing particles, there are three
different kinematical regions to consider: (i), where $u,t<0<s$;
(ii), where $u,s<0<t$; and (iii), where $s,t<0<u$. In all regions,
we have
\begin{equation}
B_7^{NP} =
{\left(i \pi^{d/2}\right)}^2 \,\Gamma^2(1+\epsilon)
 \left\{ \frac{F_t}{s^2 t} + \frac{F_u}{s^2 u} \right\}
\, .
\end{equation}
We will use the abbreviations
$T=\log(-t/s)$,
$U=\log(-u/s)$,
$V=\log(-u/t)$.

In region (i), with $u,t<0$, and $s=-t-u$, $F_t$ is given by
\begin{eqnarray}
F_t & = & s^{-2\epsilon} \left\{ \vphantom{\frac{1}{1}} \right.
\frac{-2}{\epsilon^4}
+ \frac{1}{2 \epsilon^3}
  \left( 5\, T  + 7\, U  \right)
+ \frac{1}{\epsilon^2}
  \left( 6\, U  - T^2  - 4\, T  U - U^2 
   -\frac{5}{12} \pi^2 \right)
\nonumber \\ &&
+ \frac{1}{\epsilon}
  \left( -24\, U  -12\, T  U 
     -\frac{1}{3} T^3  + 3\, T  U^2 
     - U^3 
     +\frac{1}{6} \pi^2 \left(5\, T -29\, U \right)
\right. \nonumber \\ && \left.
\hspace{2em}
     -2\, T \,\Li{2}{-t/s}
     +2\,\Li{3}{-t/s}-2\,\snp{1,2}{-t/s}
     +\frac{19}{2}\zeta(3)
  \right)
\nonumber \\ &&
 +96\, U +48\, T  U -12\, T^2  U 
 +12\, T  U^2 
 -4\, U^3 +\frac{2}{3}  T^4  +\frac{8}{3} T^3  U 
 - T^2  U^2 
\nonumber \\ &&
 -\frac{4}{3} T  U^3  +\frac{4}{3} U^4 
 + \pi^2 \left(-14\, U -\frac{5}{6} T^2 
      +\frac{38}{3} T  U 
      +\frac{25}{6} U^2 
    \right)
 +\frac{37}{40} \pi^4
\nonumber \\ &&
 -\left(13\, T +33\, U \right)\,\zeta(3)
 + \left(-48\, T +17\, T^2 
    +12\, T  U +10\,\pi^2\right)\,\Li{2}{-t/s}
\nonumber \\ &&
 + \left(48-60\, T -12\, U \right)\,\Li{3}{-t/s}
 + \left(-24+28\, T 
     -6\, U \right)\,\snp{1,2}{-t/s}
\nonumber \\ &&
 +86\,\Li{4}{-t/s}-26\,\snp{1,3}{-t/s}-36\,\snp{2,2}{-t/s}
\nonumber \\ &&
+ i\pi \left[
 \frac{2}{\epsilon^3}
 +\frac{1}{\epsilon^2}\left(6- T + U \right)
 -\frac{1}{\epsilon}\left(24+12\, T +2\, T^2 
    +2\, T  U +2\, U^2 +\frac{31}{6}\pi^2\right)
\right. \nonumber \\ &&
\hspace{2em}
 +96 +48\, T +12\, T^2 -24\, T  U 
 +\frac{10}{3} T^3 
 +2\, U^3 
\nonumber \\ &&
\hspace{2em}
  +\pi^2 \left(-2 +\frac{28}{3} T -\frac{1}{3} U \right)
 +\left(-24+14\, T  +18\, U \right)\,\Li{2}{-t/s}
\nonumber \\ &&
\hspace{2em}
 -32\,\Li{3}{-t/s} +44\,\snp{1,2}{-t/s} -89\,\zeta(3)
\left. \left. \vphantom{\frac{1}{1}} \right] \right\} \, ,
\end{eqnarray}
and $F_u$ can be obtained by interchanging $(t\leftrightarrow u)$ in
the expression for $F_t$.

In region (ii), with $u,s<0$, and $t=-s-u$, $F_t$ and $F_u$ are
given by
\begin{eqnarray}
F_t & = & t^{-2\epsilon} \left\{ \vphantom{\frac{1}{1}} \right.
\frac{-2}{\epsilon^4}
+ \frac{1}{\epsilon^3} \left( 2\,T + \frac{7}{2} V \right)
+ \frac{1}{\epsilon^2} \left( 6\,T+6\,V+2\,T^2+T V -V^2
   + \frac{31}{12} \pi^2 \right)
\nonumber \\ &&
+ \frac{1}{\epsilon} \left( -24\,T-24\,V-\frac{2}{3} T^3 -2\,T^2 V
   -2\,T V^2 -V^3
 -\frac{1}{6} \pi^2 \left(23\,T+41\,V \right)
\right. \nonumber \\ && \left.
\hspace{4em}
 +2\,\snp{1,2}{-s/t} + \frac{15}{2} \zeta(3) \right)
\nonumber \\ &&
 +96\,T + 96\,V -4\,T^3 -12\,T^2 V-4\,V^3
 -\frac{11}{6} T^4 -\frac{13}{3} T^3 V + T^2 V^2 +2\,T V^3
 +\frac{4}{3} V^4
\nonumber \\ &&
 + \pi^2 \left( -14\,T-14\,V-\frac{22}{3} T^2
 -\frac{23}{3} T V + \frac{37}{6} V^2 \right)
 -\frac{311}{120} \pi^4
\nonumber \\ &&
 -\left(24+45\,T+39\,V\right)\, \zeta(3)
 +\left(24\,T +6\,T^2-18\,T V+ 13 \pi^2 \right) \, \Li{2}{-s/t}
\nonumber \\ &&
 + \left(24+12\,T-18\,V \right) \,\Li{3}{-s/t}
 + \left(24 - 44\,T + 6\,V \right) \,\snp{1,2}{-s/t}
\nonumber \\ &&
 + 12\,\Li{4}{-s/t} + 26\,\snp{1,3}{-s/t} -62\,\snp{2,2}{-s/t}
\nonumber \\ &&
 + i \pi \left[ \frac{-5}{2 \epsilon^3}
 + \frac{1}{\epsilon^2} \left( T + 4\,V \right)
\right. \nonumber \\ &&
\hspace{2em}
 + \frac{1}{\epsilon} \left( 12\,T + 12\,V + 4\,T^2 + 2\, T V -3\, V^2
 + \frac{5}{2} \pi^2 - 2\,\Li{2}{-s/t} \right)
\nonumber \\ &&
\hspace{2em}
 -48\,T -48\,V +12\,T^2 +24\,T V -12\,V^2
 +\frac{11}{3} T^3 +2\,T^2 V + \frac{4}{3} V^3
\nonumber \\ &&
\hspace{2em}
 + \pi^2 \left( 4- 6\,V \right)
 + \left( -48 + 14\,T +12\, V \right) \Li{2}{-s/t}
\nonumber \\ && \left. \left.
\hspace{2em}
 + 32\,\Li{3}{-s/t} + 28\,\snp{1,2}{-s/t} -15\,\zeta(3)
\vphantom{\frac{1}{1}}
\right] \right\} 
\, ,
\end{eqnarray}
\begin{eqnarray}
F_u & = & t^{-2\epsilon} \left\{ \vphantom{\frac{1}{1}} \right.
\frac{-2}{\epsilon^4}
+ \frac{1}{\epsilon^3} \left( 2\,T + \frac{5}{2} V \right)
+ \frac{1}{\epsilon^2} \left( 6\,T + 2\,T^2 -T V -V^2
 + \frac{31}{12} \pi^2 \right)
\nonumber \\ &&
+ \frac{1}{\epsilon} \left( -24\,T - 12\,T V -\frac{2}{3} T^3
   - 4\,T^2 V -2\,T V^2 -\frac{1}{3} V^3
 -\frac{1}{6} \pi^2 \left( 23\,T + 31\,V \right)
\right. \nonumber \\ && \left.
\hspace{2em}
 +2\,\snp{1,2}{-s/t} + \frac{15}{2} \zeta(3)
\right)
\nonumber \\ &&
 + 96\,T + 48\,T V -4\,T^3 + 12\, T V^2
 -\frac{11}{6} T^4 - \frac{5}{3} T^3 V + 4\, T^2 V^2
 + \frac{10}{3} T V^3 + \frac{2}{3} V^4
\nonumber \\ &&
 + \pi^2 \left( -14\,T -\frac{22}{3} T^2 -\frac{7}{3} T V
    + \frac{31}{6} V^2 \right) - \frac{311}{120} \pi^4
 - \left(24 + 45\, T + 21\,V \right) \,\zeta(3)
\nonumber \\ &&
 + \left( -24\, T -6\,T^2 -18\,T V + 11 \pi^2 \right) \,\Li{2}{-s/t}
 - \left( 24 + 12\,T + 18\,V \right) \,\Li{3}{-s/t}
\nonumber \\ &&
 + \left( 48 - 32\, T + 26\,V \right) \,\snp{1,2}{-s/t}
 - 12\,\Li{4}{-s/t} + 86\,\snp{1,3}{-s/t} - 50\,\snp{2,2}{-s/t}
\nonumber \\ &&
 + i \pi \left[ \frac{-7}{2 \epsilon^3}
 + \frac{1}{\epsilon^2} \left( -6 -T + 4\,V \right)
\right. \nonumber \\ &&
\hspace{2em}
 + \frac{1}{\epsilon} \left( 24 + 12\, V + 2\,T^2 + 2\,T V - V^2
 + \frac{11}{2} \pi^2 - 2\,\Li{2}{-s/t} \right)
\nonumber \\ &&
\hspace{2em}
 -96 - 48\,V -12\,V^2 + \frac{7}{3} T^3 -4\,T^2 V -2\,T V^2 -\frac{4}{3} V^3
\nonumber \\ &&
\hspace{2em}
 + \pi^2 \left( 6 + 6\,T - 8\,V \right)
 + \left( -24+ 26\,T - 8\,V \right) \,\Li{2}{-s/t}
\nonumber \\ && \left. \left.
\hspace{2em}
 + 44\,\Li{3}{-s/t} - 24\,\snp{1,2}{-s/t} + 75\,\zeta(3)
\vphantom{\frac{1}{1}}
\right] \right\} 
\, .
\end{eqnarray}
The formulas for $F_t$ and $F_u$ in region (iii) can be obtained easily
from the ones for region (ii) by using the fact the $B_7^{NP}$ is symmetric
under $(t\leftrightarrow u)$. This follows from the symmetry between the
legs of the non-planar double box carrying momenta $p_3$ and $p_4$.

\vspace{1ex}
{\bf 6. Six propagator diagram}\\
The non-planar diagram with six propagators,
\begin{equation}
\label{eq:B6NP}
B_6^{NP} =
\int \int \mbox{d}^d k  \; \mbox{d}^d l \;
\frac{1}{(k+l+p_2)^2 (k+l)^2 (k-p_3)^2 k^2 (l-p_4)^2 l^2 }
\, ,
\end{equation}
is far less difficult to calculate than $B_7^{NP}$. Perhaps the easiest
way is by considering a scale transformation,
\begin{equation}
\label{eq:b6scaletrf}
B_6^{NP}(\lambda p_1, \lambda p_2, \lambda p_3, \lambda p_4) =
\lambda^{2(d-6)} B_6^{NP}(p_1, p_2, p_3, p_4) \, .
\end{equation}
Differentiating both sides of eq.~(\ref{eq:b6scaletrf}) with respect
to $\lambda$ at $\lambda=1$ gives an identity which can be used to 
reduce $B_6^{NP}$ to a sum of three planar diagrams:
\begin{equation}
\label{eq:b6-deco}
\begin{minipage}{3cm}
\centering
\includegraphics[height=1.5cm]{b6c.eps}
\end{minipage}
= \frac{-1}{1+4 \epsilon}
\left\{ \;
\begin{minipage}{3cm}
\centering
\includegraphics[height=1.5cm]{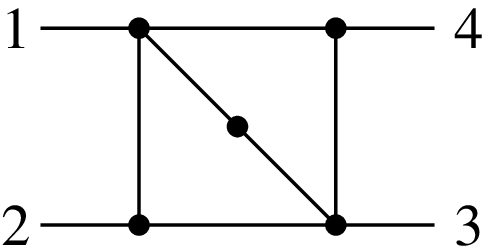}
\end{minipage}
+
\begin{minipage}{3cm}
\centering
\includegraphics[height=1.5cm]{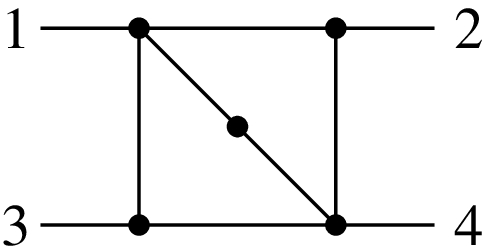}
\end{minipage}
+
\begin{minipage}{3cm}
\centering
\includegraphics[height=1.5cm]{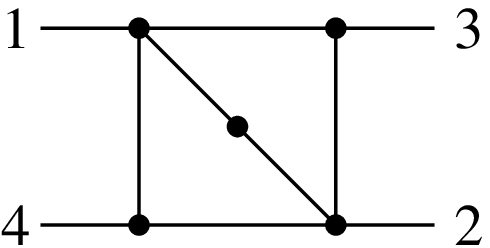}
\end{minipage}
\; \right\} \, .
\end{equation}
In the diagrams on the right hand side of eq.~(\ref{eq:b6-deco}),
the propagator on the diagonal is squared. Such boxes with a diagonal
propagator are calculated in ref.~\cite{SmirnovVeretin99}. They are,
in fact, rather similar to one-loop massless box diagrams~\cite{Willy86}.
The pole at $\epsilon=-1/4$ is a reflection of the linear infrared
divergence\footnote{It would diverge as $1/m_{reg}$ if we used a
mass $m_{reg}$ to regularize it.} in (\ref{eq:B6NP}) coming from the
region where the loop momenta $k$ and $l$ are both soft. Another curious
feature of eq.~(\ref{eq:b6-deco}) is that although the three terms on the
right hand side contain third order poles at $\epsilon=0$, the leading
singularity of the sum is only $1/\epsilon^2$.

The six propagator diagram is completely symmetric under permutations
of its external momenta. Therefore, it is sufficient to consider only
region (i), where $u,t<0<s$ and $s=-t-u$. We obtain
\begin{eqnarray}
B_6^{NP} \!\! & = \!\! & {(i \pi^{d/2})}^2
\frac{\Gamma(1+2 \epsilon) \Gamma^3(1-\epsilon)}
{\Gamma(1-3 \epsilon)(1+4 \epsilon)}
\frac{3}{\epsilon^2}
\left\{  \frac{s^{-2\epsilon}}{s\,u}
\left(\vphantom{\frac{1}{1}}
  \left( T + i \pi \right) \left(1-2\epsilon U\right)
\right. \right. \nonumber \\ &&
+ 2 \epsilon^2 \left(
  2\, \Li{3}{-t/s} - 4\, \snp{1,2}{-t/s}
 + 2\, \zeta(3) - \frac{1}{3} T^3
\right. \nonumber \\ && \left. \left.
  + T \left[ -2\, \Li{2}{-t/s} + U^2 - \pi^2 \right]
  + i \pi \left[ 2\, \Li{2}{-t/s} + U^2 -\frac{1}{3} \pi^2 \right]
\right) \right) \nonumber \\ &&
\left. \vphantom{\frac{1}{1}} + (u \leftrightarrow t) \right\} \, .
\end{eqnarray}

The reduction formula~(\ref{eq:b6-deco}) can be generalized to the case
where the six propagators in eq.~(\ref{eq:B6NP}) are raised to arbitrary
integer powers.

\vspace{1ex}
{\bf 7. Discussion}\\
The analytical property that distinguishes the
non-planar diagrams from the planar ones considered in
refs.~\cite{Smirnovdoublebox,SmirnovVeretin99,AGO99ii} is the fact
that they have cuts in three channels, $s$, $t$, and $u$, rather than
just in two. As a consequence, they have imaginary parts in all three
physical regions, (i), (ii) and (iii). This is why, when we derived the
Mellin-Barnes representation~(\ref{eq:XIMB}) for the seven propagator
diagram, we were led to considering it as a function of three independent
variables. A priori, one would think that going from a two-scale problem
to a three-scale problem could make it vastly more difficult to solve,
but it did not, because, thanks to the vanishing of~(\ref{eq:zero}),
all genuinely three-scale contributions cancelled out.

Generalizing the Mellin-Barnes representation for the seven
propagator diagram to arbitrary powers of propagators is completely
straightforward, and merely amounts to shifting the arguments of the
$\Gamma$-functions~(\ref{eq:gammalist}) by some constants. It is likely
that cancellations similar to~(\ref{eq:zero}) will take place for other
integer powers, so that the method described in this paper could be used
for those cases as well.

\vspace{1ex}
I am very grateful to T.~Binoth and G.~Heinrich for their help with
the numerical checks, and for many discussions.  I would also like
to thank J.J.~van~der~Bij, A.I.~Davydychev, T.~Gehrmann, K.~Melnikov,
E.~Remiddi and V.A.~Smirnov for helpful discussions and correspondence.
This work was supported by the DFG-Forschergruppe ``Quantenfeldtheorie,
Computeralgebra und Monte-Carlo-Simulation''.

\end{document}